\documentclass{ws-ijqi}

\newcommand{\DD}{\mathit{\Delta}}
\begin{document}

\markboth{Robabeh Rahimi, Akira SaiToh, Mikio Nakahara}
{Coherence Conservation of a Qubit Coupled to a Thermal Dissipating Environment}


\title{COHERENCE CONSERVATION OF A QUBIT\\ 
COUPLED TO A DISSIPATING THERMAL ENVIRONMENT}

\author{ROBABEH RAHIMI}

\address{Interdisciplinary Graduate School of Science and Engineering, Kinki University,\\
Higashi-Osaka, Osaka 577-8502, Japan\\
rahimi@alice.math.kindai.ac.jp}

\author{AKIRA SAITOH}

\address{Department of Systems Innovations, Graduate School of Engineering Science, Osaka University\\
Toyonaka, Osaka, 560-8531, Japan\\
saitoh@qc.ee.es.osaka-u.ac.jp}

\author{MIKIO NAKAHARA}

\address{Department of Physics, Kinki University\\
Higashi-Osaka, Osaka 577-8502, Japan\\
nakahara@math.kindai.ac.jp}

\maketitle


\begin{abstract}
It is shown that quantum coherence is conserved in a principal system in the case that the system is coupled to a fast dissipating environment [arXiv:0709.0562]. The phenomenon is called the quantum wipe effect. Here, this effect is reviewed and the analytical proof for a model system consisting of a  one-qubit system coupled to a fast dissipating environment is extended to an environment at a thermal equilibrium. 
\end{abstract}


\section{Introduction}
A conventional approach in facing the problem of decoherence in a system, most usually, is by concentrating on the principal system itself. It is somehow implicitly assumed that there is no control over the environmental system \cite{1}${}^-$\cite{7}. Then, by operating with the system itself, for instance, by applying on it, fast and strong multi-pulses for a dynamical control of the system \cite{2,3}, the decoherece is aimed to be suppressed. On the other hand, the situation could be entirely different if one could have access to the environment for controlling the decoherence of the system.

From the first point of view, the idea of suppressing decoherence of a principal system by controlling the environment instead of working directly with the principal system might seem not to be reasonable. This is true specially when one tries to decrease the noisy behavior of the environment for stabilizing the principal system. However, a control of the decoherence of a system is attained by making the environment even noisier, then this approach should turn out to be feasible. In a numerical simulation of bang-bang control of entanglement in a spin-bus model, decoherence in the principal system is shown to be suppressed if the environment is made to be rapidly dissipating to a very large bath environment \cite{bb}. In fact. simillar phenomena in different models are known in the community \cite{new1,new2}.

We have investigated the concept of controlling the decoherence of a system while the corresponding environment dissipates fast into another larger environment that can be a bath system \cite{ours}. This phenomenon is called the ``quantum wipe effect''. This effect is proved analytically for the case of a one-qubit principal system when it is coupled to a maximally mixed state environment that is dissipating to a large bath environment. Numerical evaluations for a single-qubit principal system coupled to a dissipating bosonic environment is conducted in addition to the example of an entangled two spin state as a principal system coupled to the environment \cite{ours}. Here, we extend the analytical proof of the previous work to the case where the principal system is a one-qubit system coupled with a fast dissipating environment that is initially in a Boltzmann distribution rather than a simple maximally mixed state.

In the following section the model for this study is introduced. In section 3 the analytical proof is given. In section 4, findings of the mathematical proof is discussed in more details, giving an sketched overview on the phenomenon itself and explaining the conditions under which the phenomenon can be effective.

\section{Model}
We assume a model system involving a principal system (system 1) coupled to an environment (system 2). The system is represented by $\rho^{[1,2]}$. We further assume a large thermal environment system surrounding the systems 1 and 2, such that the state of the system 2 is replaced by that of the thermal environment with probability $p$ (namely, with some dissipation rate) per unit time interval $\tau$. The thermal environment is represented by the density matrix $\sigma$.

\begin{figure}[tb]
\begin{center}
 \scalebox{0.6}{\includegraphics{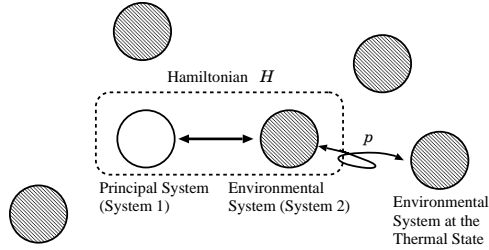}}
\caption{\label{figsimplemodel}
Model for the system consisting of the principal system (system 1) and the environmental system (system 2) whose time evolution is governed by the  Hamiltonian $H$. System 2 is replaced with a thermal environmental system with the dissipation probability $p$ for the time interval $\tau$.}
\end{center}
\end{figure}

The Hamiltonian affecting the time evolution is reduced to the one consisting only of the time-independent Hamiltonian $H$ that governs systems 1 and 2 including their interaction. This model is illustrated in Fig.\ \ref{figsimplemodel}. For a small time interval $\DD t$, the evolution of the systems 1 and 2 obeys the equation
\begin{equation}\label{eqMain}
\rho^{[1,2]}(\tilde{t}+\DD t)=e^{-iH\DD t}\biggl[x^{\DD t}\rho^{[1,2]}(\tilde{t})+(1-x^{\DD t}){\rm Tr}_2\rho^{[1,2]}(\tilde{t})\otimes \sigma\biggr]e^{iH\DD t},
\end{equation}
where $x=(1-p)^{1/\tau}$ and $\tilde{t}$ denotes a certain time step.
The dissipation rate $p$ can be modified by changing the experimental setup under a static control.

For this model it is shown \cite{ours} that coherence conservation is achieved by $p$ very close to $1$, meaning that the environmental system 2 is rapidly dissipating to the thermal environment. For the case that the principal system is a one-qubit system and the system 2 is a maximally mixed state, the analytical proof is previously given, in addition to numerical evaluations of the effect for other cases of which the principal system is a two-qubit system or an entanglement of two spins \cite{ours}. Here, we extend the analytical proof to a more operationally reasonable case where the system 2 is in the Boltzmann thermal state.

\section{Qubit-qubit coupling with a thermal environmental qubit}
Let us consider the following setting. The principal system is originally represented by a density matrix
\begin{equation}\label{EqInitial}
 \rho^{[1]}(0)=\begin{pmatrix}a&b\\b^*&1-a\end{pmatrix}
\end{equation}
with $0\le a \le 1$ and $0\le |b|\le \sqrt{a(1-a)}$. The environmental system at thermal equilibrium is represented by the thermal density matrix (under the high temperature approximation)
\[
 \rho^{[2]}(0)=\sigma=\begin{pmatrix}(1+\epsilon)/2&0\\
 0&(1-\epsilon)/2\end{pmatrix}
\]
with $\epsilon=\tanh[E_\Delta/(2k_{\rm B}T)]$, the polarization for the Zeeman energy $E_\Delta$ and temperature $T$ ($k_{\rm B}$ is the Boltzmann constant). The initial state of the total system is set to
$
 \rho^{[1,2]}(0)=\rho^{[1]}(0)\otimes \rho^{[2]}(0).
$
The Hamiltonian $H$ is set to $cI_z\otimes I_z=\mathrm{diag}(c/4,-c/4,-c/4,c/4)$ [here, $I_z=\mathrm{diag}(1/2, -1/2)$].

Under these conditions, $\rho^{[1,2]}$ at time $t=m\DD t$ ($m\in\{0,1,2,\ldots\}$) is given as
\[
\rho^{[1,2]}(m\DD t)=
 \begin{pmatrix}
 a(1+\epsilon)/2&0&f_m&0\\
 0&a(1-\epsilon)/2&0&g_m\\
 f^*_m&0&(1-a)(1+\epsilon)/2&0\\
 0&g^*_m&0&(1-a)(1-\epsilon)/2
 \end{pmatrix},
\]
with functions $f_m$ and $g_m$ depending on $m$, satisfying the system of recurrence formulae as follows
\[
\left\{
\begin{array}{l}
f_{m+1}=e^{-ic\DD t/2}\left[x^{\DD t}f_m+\frac{1+\epsilon}{2}(1-x^{\DD t})(f_m+g_m)\right]\\
g_{m+1}=e^{ic\DD t/2}\left[x^{\DD t}g_m+\frac{1-\epsilon}{2}(1-x^{\DD t})(f_m+g_m)\right]
\end{array}
\right.
\]
with $f_0=b(1+\epsilon)/2$ and $g_0=b(1-\epsilon)/2$. This leads to the following recurrence formula:
\begin{equation}
\label{EqRec}\begin{split}
&\kappa_{m+2}-\kappa_{m+1}\left\{\left(\frac{1+\epsilon}{2}+\frac{1-\epsilon}{2}x^{\DD t}\right)e^{-ic\DD t/2}+\left(\frac{1-\epsilon}{2}+\frac{1+\epsilon}{2}x^{\DD t}\right)e^{ic\DD t/2}\right\}\\
&+ \kappa_m x^{\DD t}=0,
\end{split}
\end{equation}
where $\kappa_m=f_m$ or $g_m$ with $f_0$ and $g_0$ as introduced above, and $f_1=b(1+\epsilon)e^{-ic\DD t/2}/2$, and $g_1=b(1-\epsilon)e^{ic\DD t/2}/2$.

One can derive functions $f(t)={\rm lim}_{\DD t\rightarrow 0, m\DD t = t}f_m$ and $g(t)={\rm lim}_{\DD t\rightarrow 0, m\DD t = t}g_m$ in the following way. By linearization, Eq.\ (\ref{EqRec}) is put in the form:
\[
\begin{split}
&\kappa_{m+2}-2\kappa_{m+1}+\kappa_m-\DD t\ln x(\kappa_{m+1}-\kappa_m)+\frac{(\DD t)^2}{2}(c^2/2-ic\epsilon\ln x)\kappa_{m+1}\\
&-\frac{(\DD t)^2}{2}(\ln x)^2(\kappa_{m+1}-\kappa_m)+\mathcal{O}[(\DD t)^3]=0.
\end{split}\]
Dividing this equation by $(\DD t)^2$ and taking the limit $\DD t\rightarrow 0$ lead to
\[
 \frac{\partial^2}{\partial t^2}\kappa(t)-\ln x ~\frac{\partial}{\partial t}\kappa(t)+\left(\frac{c^2}{4}-\frac{ic\epsilon\ln x}{2}\right)\kappa(t)=0,
\]
where $\kappa(t)={\rm lim}_{\DD t\rightarrow 0, m\DD t = t}\kappa_m$. The solution of this differential equation is
\[
 \kappa(t)=u_\kappa e^{-r_+ t} + v_\kappa e^{-r_- t}
\]
with constants $u_\kappa$ and $v_\kappa$ ($\kappa=f$ or $g$), and the complex decoherence factor
\begin{equation*}\label{eqr}
r_\pm=-\frac{1}{2}\left[\ln x\pm\sqrt{(\ln x)^2-c^2+i2c\epsilon\ln x}\right].
\end{equation*}

We need to impose the conditions that $f(0)= b(1+\epsilon)/2$, $g(0)=b(1-\epsilon)/2$, and $\kappa'(0)=\lim_{\DD t\rightarrow 0} (\kappa_1-\kappa_0)/\DD t$. The latter condition can be written as $-r_+u_f -r_-v_f=-ibc(1+\epsilon)/4$ and $-r_+u_g -r_-v_g=ibc(1-\epsilon)/4$, for $\kappa = f$ and $g$,
respectively. Thus we obtain
\[\begin{split}
u_f &= \frac{-b(1+\epsilon)}{2(r_+-r_-)}\left(r_--i\frac{c}{2}\right),~~v_f =  \frac{b(1+\epsilon)}{2(r_+-r_-)}\left(r_+-i\frac{c}{2}\right),\\
u_g &= \frac{-b(1-\epsilon)}{2(r_+-r_-)}\left(r_-+i\frac{c}{2}\right),~~v_g =  \frac{b(1-\epsilon)}{2(r_+-r_-)}\left(r_++i\frac{c}{2}\right). 
\end{split}\]
Consequently, we have
\[
 f(t)=\frac{b(ic-2r_-)(1+\epsilon)}{4(r_+-r_-)}e^{-r_+t}+\frac{b(-ic+2r_+)(1+\epsilon)}{4(r_+-r_-)}e^{-r_-t},
\]
\[
 g(t)=\frac{b(-ic-2r_-)(1-\epsilon)}{4(r_+-r_-)}e^{-r_+t}+\frac{b(ic+2r_+)(1-\epsilon)}{4(r_+-r_-)}e^{-r_-t}.
\]

One can now write the reduced density matrix of the principal system at $t$ as
\begin{equation}\label{EqFinal}
 \rho^{[1]}(t)=\begin{pmatrix} a& \eta(t)\\
\eta(t)^*&1-a
\end{pmatrix}
\end{equation}
with
\begin{equation}\label{Final}
 \eta(t)=b\left(\frac{-r_-+ic\epsilon/2}{r_+-r_-}e^{-r_+ t}+\frac{r_+-ic\epsilon/2}{r_+-r_-}e^{-r_- t}\right).
\end{equation}

\begin{figure}[bt]
\begin{center}
\begin{minipage}{0.49\textwidth}
 \hspace{-0.5cm}\scalebox{0.36}{\includegraphics{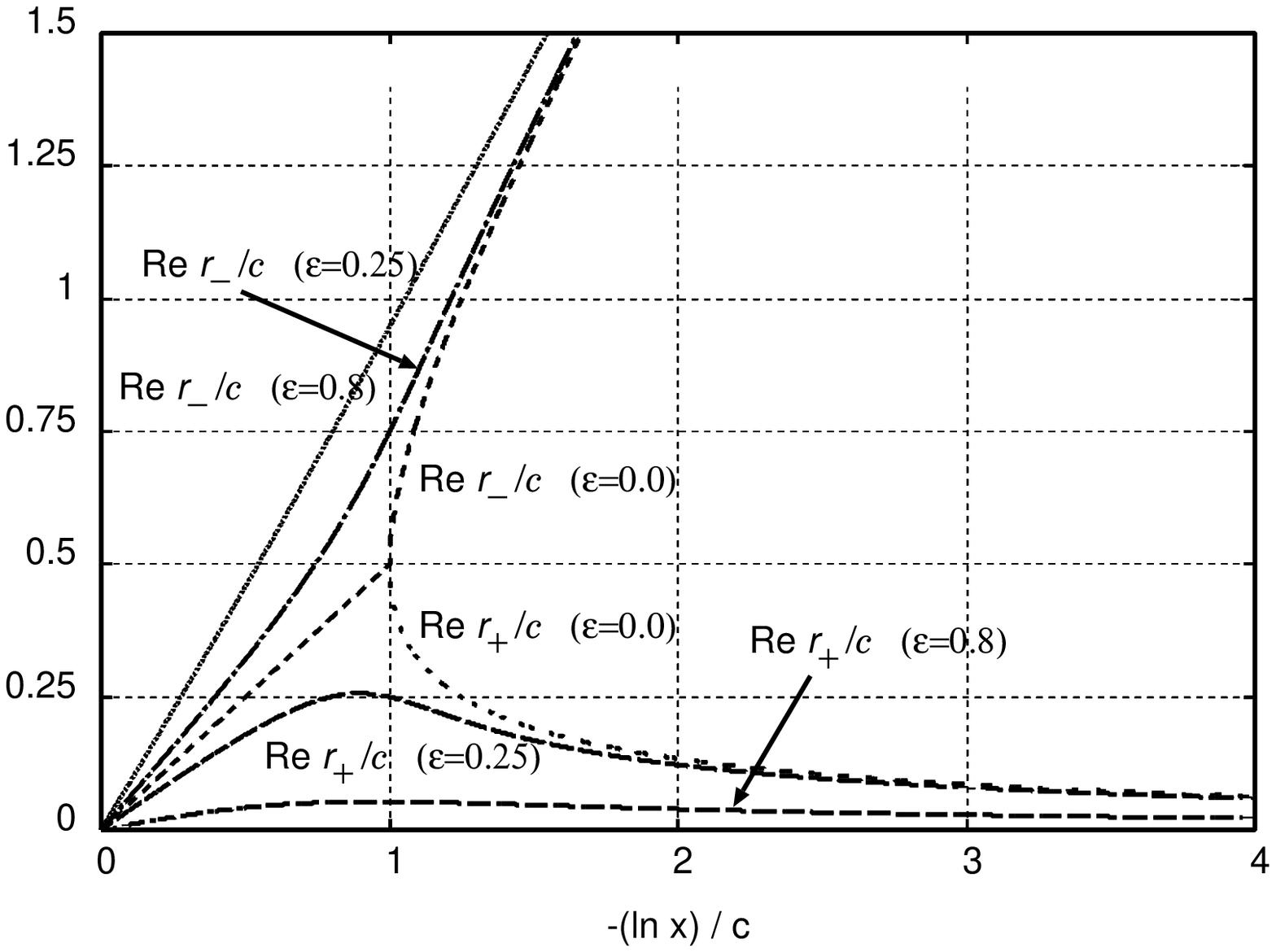}}\\
$~\hspace{0.4\textwidth}$(a)
\end{minipage}
\begin{minipage}{0.49\textwidth}
 \hspace{-0.4cm}\scalebox{0.36}{\includegraphics{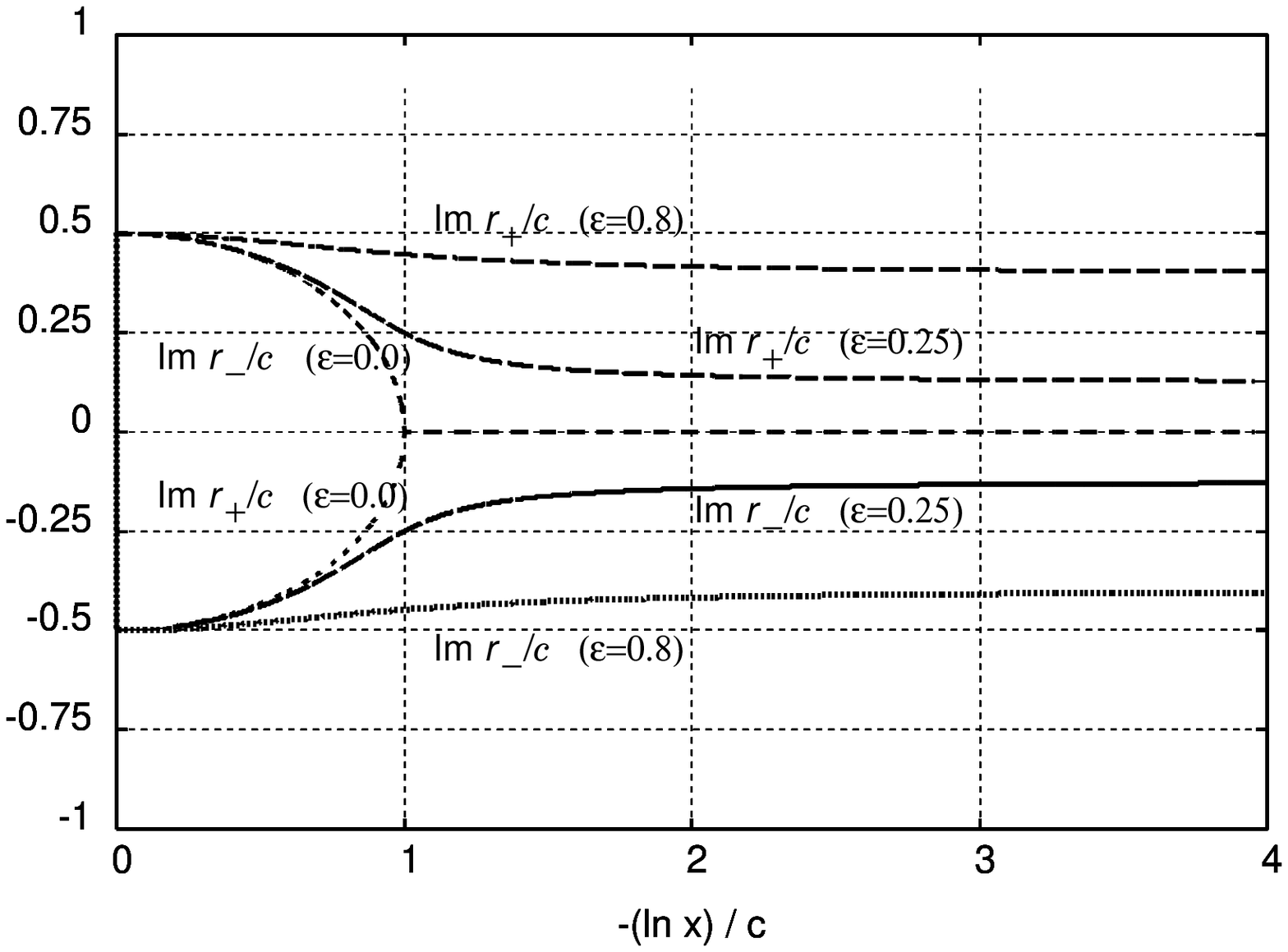}}\\
$~\hspace{0.4\textwidth}$(b)
\end{minipage}
\caption{\label{fig1}(a) Plots of ${\rm Re}~r_\pm/c$ as functions of $-(\ln x) / c$ for different values of polarization $\epsilon$. For $\epsilon=0$, the decoherence factors ${\rm Re}~r_\pm$ increases until they reach $c/2$ as the dissipation rate increases; the factor ${\rm Re}~r_+$ starts decreasing at the point of $-\ln x = c$ (i.e., $p=1-e^{-c\tau}$) while the factor ${\rm Re}~r_-$ starts increasing rapidly at this point. For other values of $\epsilon$, the behavior is similar to that of $\epsilon=0$.
(b) Plots of ${\rm Im}~r_\pm/c$ as functions of $-(\ln x) / c$ for different values of polarization $\epsilon$.}
\end{center}
\end{figure}

It is possible to realize that if $p$ increases to reach one, then $|\eta(t)|$ converges to $b$, giving the coherence of $\rho^{[1]}(t)$, Eq.\ (\ref{EqFinal}), equal to that of $\rho^{[1]}(0)$, Eq.\ (\ref{EqInitial}), ignoring an unimportant phase factor. This is clear if one investigates the behavior of $r_\pm$ in details in relation to $p$ and $\epsilon$. Figure\ \ref{fig1}, (a) and (b), shows the real and imaginary parts of $r_\pm$ as functions of $-(\ln x)/c$ for several different values of $\epsilon$. It is clear that the total behavior of ${\rm Re}~r_\pm$ does not very much depend on different values of $\epsilon$. However, by increasing $-(\ln x)/c$, namely by increasing $p$, the decoherence factors ${\rm Re}~r_\pm$ increase until they reach a certain value (e.g. $c/2$ for the case of $\epsilon=0$) then the factor ${\rm Re}~r_+$ starts decreasing while the factor ${\rm Re}~r_-$ starts increasing very rapidly. Fig.\ \ref{fig1} (b) shows that as $-(\ln x)/c$ increases, namely as $p$ approaches to unity, ${\rm Im}~r_\pm$ do not have any large change.

\begin{figure}[bt]
\begin{center}
\begin{minipage}{0.49\textwidth}
\scalebox{0.48}{\includegraphics{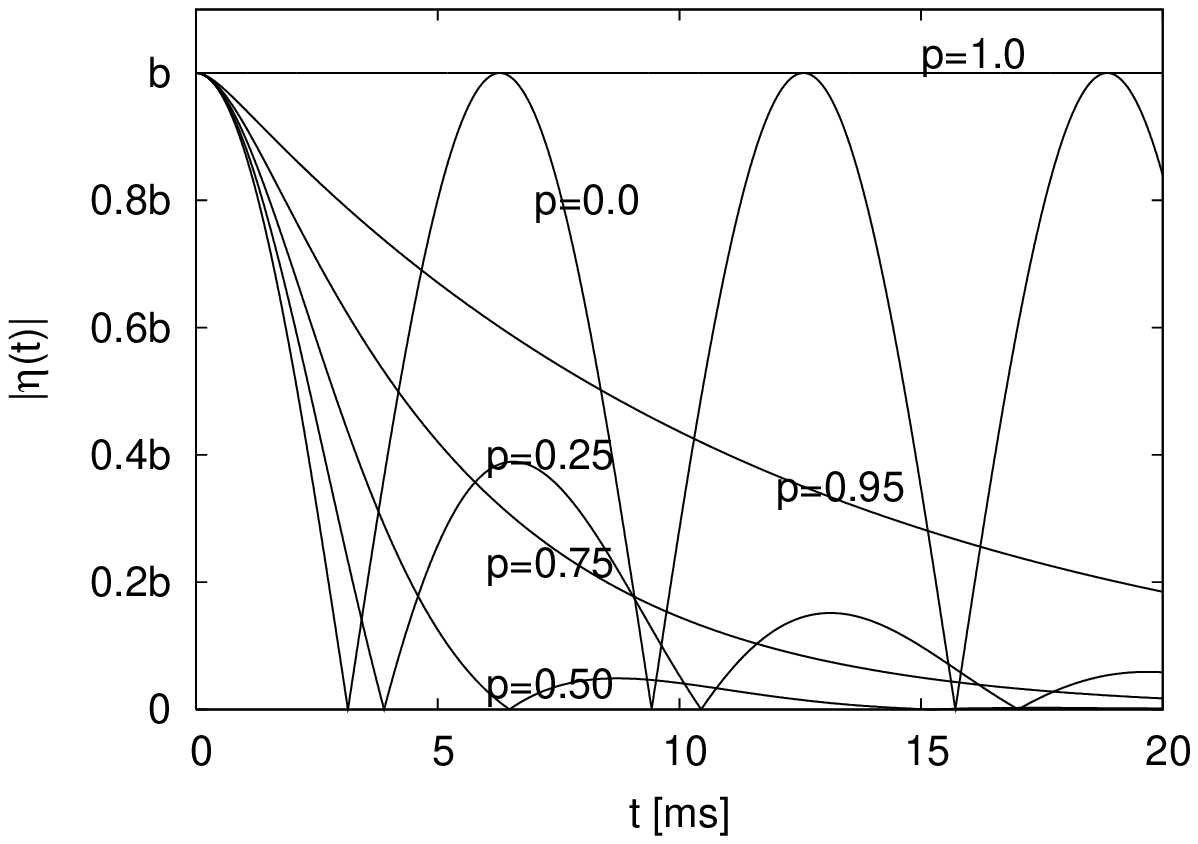}}\\
$~\hspace{0.4\textwidth}$(a)
\end{minipage}
\begin{minipage}{0.49\textwidth}
\scalebox{0.48}{\includegraphics{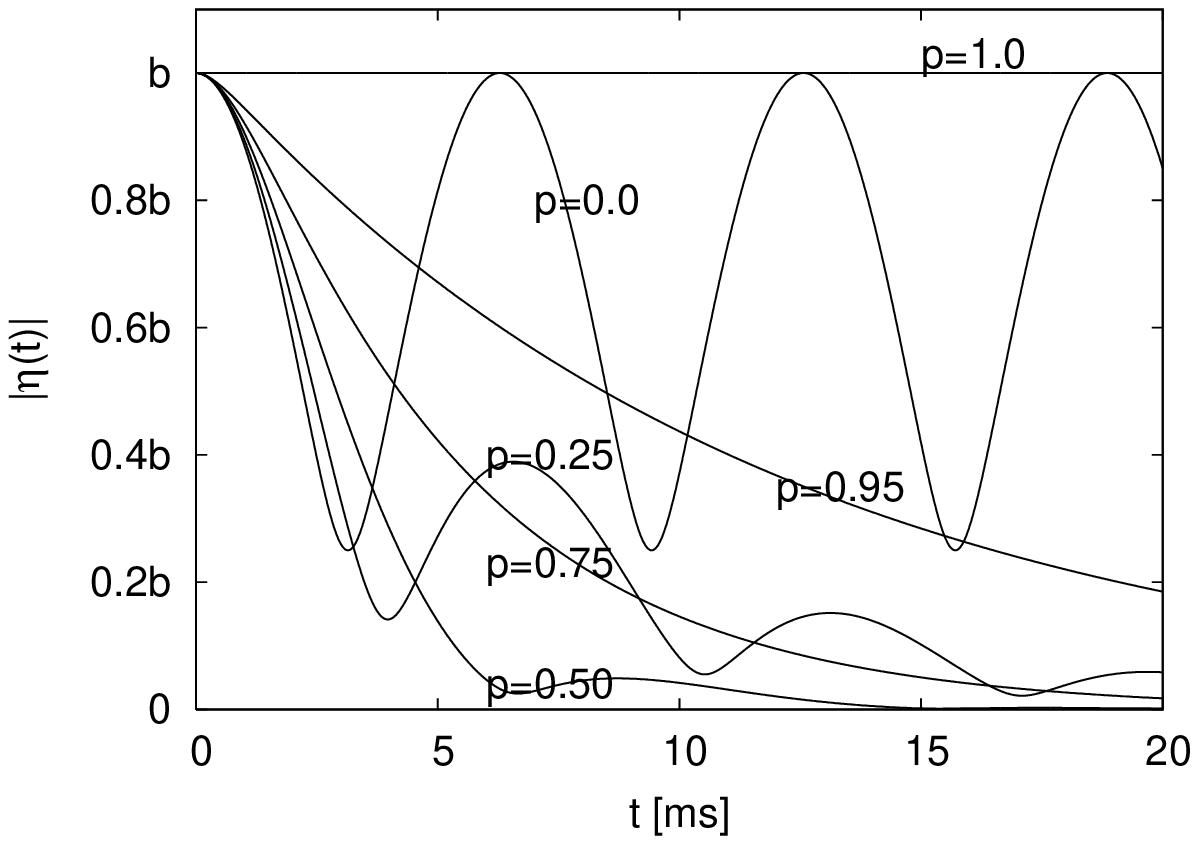}}\\
$~\hspace{0.4\textwidth}$(b)
\end{minipage}\\
\begin{minipage}{0.49\textwidth}
\scalebox{0.48}{\includegraphics{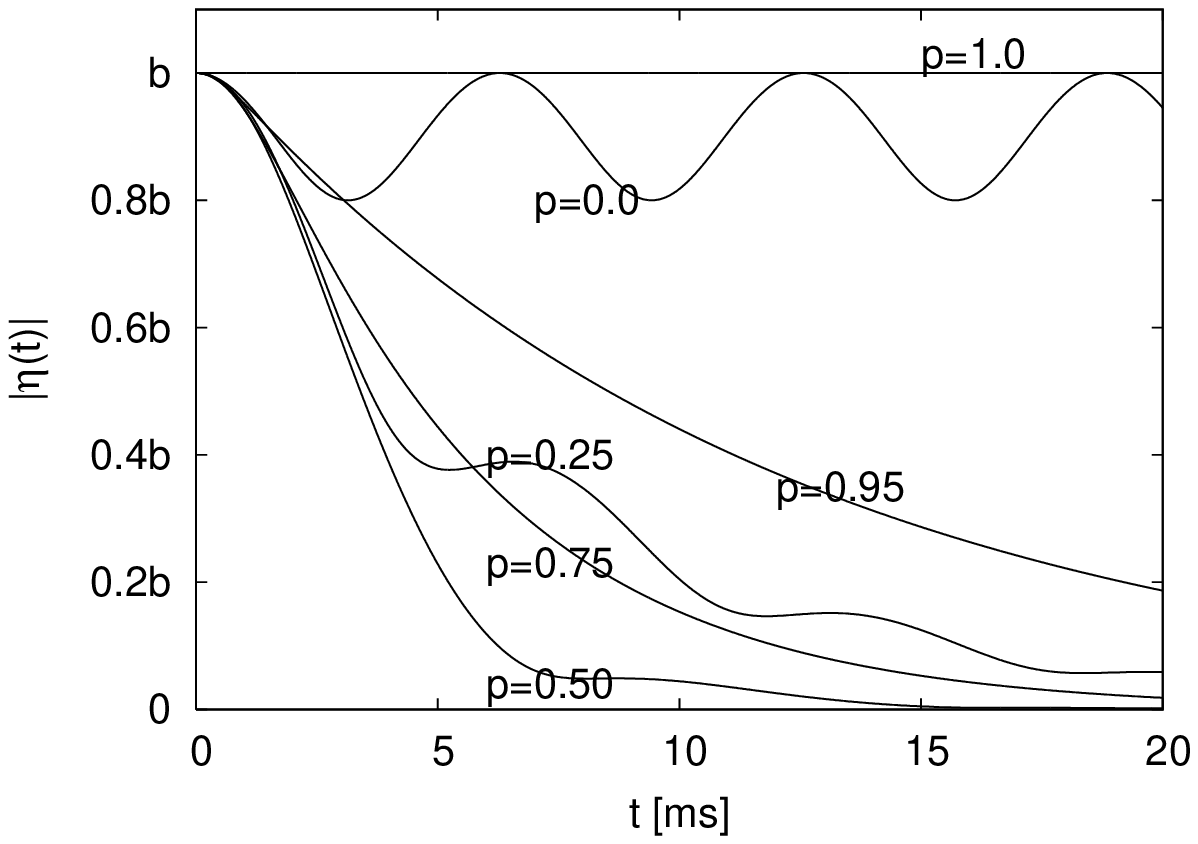}}\\
$~\hspace{0.4\textwidth}$(c)
\end{minipage}
\begin{minipage}{0.49\textwidth}
\caption{\label{fig2}(a) Time evolution of $|\eta(t)|$ for several different values of $p$ ($0.0$, $0.25$, $0.5$, $0.75$, $0.95$, and $1.0$) when $\epsilon=0.0$, $\tau=1.0\times 10^{-3}$s and $c=1.0\times 10^3$Hz.
(b) Time evolution of $|\eta(t)|$ for several different values of $p$ ($0.0$, $0.25$, $0.5$, $0.75$, $0.95$, and $1.0$) when $\epsilon=0.25$, $\tau=1.0\times 10^{-3}$s and $c=1.0\times 10^3$Hz.
(c) Time evolution of $|\eta(t)|$ for several different values of $p$ ($0.0$, $0.25$, $0.5$, $0.75$, $0.95$, and $1.0$) when $\epsilon=0.8$, $\tau=1.0\times 10^{-3}$s and $c=1.0\times 10^3$Hz.}
\end{minipage}
\end{center}
\end{figure}

These plots help us to depict the overall behavior of $\eta(t)$, Eq.\ (\ref{Final}), for $p$ close to one. The imaginary terms, for such $p$, contribute to the phase factor of $\eta(t)$ mainly which is not a significant factor of coherence. Among real factors, ${\rm Re}~r_+$ contributes to $\eta(t)$ through the first term of Eq.\ (\ref{Final}), since ${\rm Re}~r_+$ converges to zero for $p$ close to one. However, if ${\rm Re}~r_-$ becomes very large then the second term of Eq.\ (\ref{Final}) does not have a big contribution. Then $|\eta(t)|$ in the limit of $p$ close to one converges to $b$ and gives $\rho^{[1]}(t)$, Eq.\ (\ref{EqFinal}), equal to $\rho^{[1]}(0)$, Eq.\ (\ref{EqInitial}), ignoring the phase of $\eta(t)$. One can conclude that for large dissipation rate $p$, decoherence does not have effect on the principal system.
The convergence of $|\eta(t)|$ to $b$ for large dissipation rate is clearly depicted in Fig.\ \ref{fig2} in which the time evolution of $|\eta(t)|$ is shown for several different values of $p$ when $\epsilon=0.0$ (Fig.\ \ref{fig2} (a)) and $\epsilon=0.25$ (Fig.\ \ref{fig2} (b)), and $\epsilon=0.8$ (Fig.\ \ref{fig2} (c)).

\section{Conclusion}
We consider decoherence in a model that involves a one-qubit principal system coupled to an environment in the Boltzmann distribution, in which the environment itself rapidly dissipates to a large bath environment. We have shown that decoherence of the principal system is suppressed for very large dissipation rates from the Boltzmannian environmental to the large bath environment. This phenomenon is called the quantum wipe effect\cite{ours} and can be understood as follows. If the dissipation rate is very large then the environmental system does not have enough time to affect the principal system for absorbing coherence information from the principal system. Thus the environmental system is {\it wiped out} and the decoherence of the principal system is suppressed without touching the principal system. It is hoped that this effect will be investigated extensively to ease the static control of decoherence.

\section*{Acknowledgments}

RR is supported by the Grant-in-Aid for JSPS fellows (Grant No. 1907329). AS is supported by the Grant-in-Aid for JSPS fellows (Grant No. 1808962). MN would like to thank for partial support of the Grant-in-Aid for Scientific Research from JSPS (Grant No. 19540422). This work is partially supported by ``Open Research Center" Project for Private Universities: matching fund subsidy from MEXT.

\end{document}